# AUTOMATED DEBUGGING IN JAVA USING OCL AND JDI [*]

David J. Murray
Lehigh University
Bethlehem, PA 18015
dama@lehigh.edu

Dale E. Parson
Bell Labs, Lucent Technologies
Allentown, PA 18013
dparson@lucent.com

## Abstract

Correctness constraints provide a foundation for automated debugging within object-oriented systems. This paper discusses a new approach to incorporating correctness constraints into Java development environments. Our approach uses the Object Constraint Language ("OCL") as a specification language and the Java Debug Interface ("JDI") as a verification API. OCL provides a standard language for expressing object-oriented constraints that can integrate with Unified Modeling Language ("UML") software models. JDI provides a standard Java API capable of supporting type-safe and side effect free runtime constraint evaluation. The resulting correctness constraint mechanism: (1) entails no programming language modifications; (2) requires neither access nor changes to existing source code; and (3) works with standard off-the-shelf Java virtual machines ("VMs"). A prototype correctness constraint auditor is presented to demonstrate the utility of this mechanism for purposes of automated debugging.

## 1 Introduction

This work arises out of the RTEEM project at Lucent Technologies. RTEEM is a research project relating to the development of a next generation successor to the LUxWORKS family of software tools for embedded system architectures [1, 2]. Within the context of RTEEM, correctness constraints — in the form of axiomatic assertions — serve two distinct purposes: (1) they provide a means for specifying and validating the semantics of RTEEM project components; and (2) they address the need for high level language debugging facilities for embedded system architectures.

The use of axiomatic assertions to specify the semantics of programming language constructs is hardly a new idea. Pioneering work by Hoare, among others, demonstrated the utility of axiomatic specification for purposes of reasoning about program correctness [3, 4]. Ultimately, Meyer's less formal *Design by Contract* methodology paved the way for widespread use of axiomatic assertions in object-oriented systems. As embodied in Eiffel, such assertions can be used not only for purposes of specification, but also for purposes of runtime validation [5, 6, 7].

The *Design by Contract* methodology employs three different types of assertions to express correctness constraints for object-oriented programs. Preconditions express properties that must be true on method entry to ensure programmatically correct execution. Postconditions express properties that must be true upon method exit, assuming that the preconditions have been met. Class invariants express properties that must be true for class objects when they are in an observable state (*i.e.*, after object creation and before and after every feature call). In a nutshell, *Design by Contract* states that if the preconditions are met, the postconditions will be guaranteed and the class invariants will be preserved. As a result, *Design by Contract* provides a solid mathematical foundation for reasoning about the correctness of object-oriented programs.

Design by Contract also provides a powerful model for object-oriented debugging. Under this model, clients are responsible for ensuring preconditions and servers are responsible for guaranteeing postconditions and preserving invariants. Accordingly, when a precondition is violated, the bug may be attributed to client code. Similarly, when a postcondition or invariant is violated, the bug may be attributed to server code. With runtime assertion monitoring, the whole process of bug attribution becomes fully automated.

---

In light of these substantial benefits, it is not surprising that correctness constraints rank number two among the requests for enhancement to the Java Programming Language [8]. Since RTEEM envisions the use of Java as an implementation language for certain of its components, we decided to explore the feasibility of incorporating correctness constraints into Java programs. We looked at a number of existing tools for this purpose, but all had certain drawbacks. Therefore, we adopted a new approach to this issue.

Our approach uses OCL as a specification language and JDI as a verification API. OCL provides a standard language for expressing object-oriented constraints with the added benefit of UML model integration. JDI provides a standard Java API capable of supporting type-safe and side effect free runtime constraint evaluation. The resulting correctness constraint mechanism: (1) entails no programming language modifications; (2) requires neither access nor changes to existing source code; and (3) works with standard off-the-shelf Java VMs.

## 2   Object Constraint Language

In November 1997, the Object Management Group ("OMG") adopted UML as a standard for object-oriented design and analysis. UML is a graphical language that can be used to "visualize, specify, construct and document the artifacts of a software-intensive system." [9, 10]  Within UML, OCL is the standard expression language for specifying preconditions, postconditions, invariants and other constraints [11, 12]. We decided to use OCL as our expression language for the following reasons: (1) it represents an industry standard; (2) it offers UML model integration; (3) it provides a relatively informal expression syntax that is more expressive than the one found in Eiffel; and (4) most of its syntax maps to Java in a relatively straightforward way.

OCL is a declarative language. Therefore, OCL expressions are guaranteed to be side effect free. OCL is also a typed language. Therefore, every OCL expression has a type. OCL includes a number of basic and collection types, and supports user defined types as well. The basic types — which include: OclAny, OclType, Real, Integer, Boolean and String — can be mapped to basic Java types (*e.g.*, java.lang.Object, java.lang.Class, float, int, boolean and java.lang.String). The collection types — which include: Set, Bag and Sequence — can be mapped to Java collection types (*e.g.*, java.util.Set and java.util.List). Although these mappings are not entirely one-to-one, they are sufficient for purposes of expressing Java-based OCL constraints.

OCL constraints are OCL expressions of type Boolean. In addition, every OCL constraint has a particular context. The context of an invariant is a class, interface or type. The context of a pre- or postcondition is an operation or a method. Explicit references to the context object are made using the keyword *self*. Additional keywords include *result* and the *@pre* suffix, which are used to express postcondition constraints. Furthermore, OCL supports first order predicate logic statements with universal and existential quantification operators. As a result, OCL constraints are more expressive than Eiffel's assertions.

```
class BoundedStack extends Object{

   private Vector v = null;
   public Stack(int size){v = new Vector(size);}
   public Object push(Object obj){
      v.add(obj);
      return v.lastElement();
   }
   public Object pop(){return v.remove(v.size()-1);}
   public Object peek(){return v.lastElement();}
   public boolean empty(){return (v.size() == 0);}
   public int size(){return v.size();}
   private int capacity(){return v.capacity();}
}
```

**Figure 1: BoundedStack Class**

As a simple example of how OCL constraints can be used in connection with Java programs, consider the BoundedStack class set forth in Figure 1. BoundedStack objects represent a common LIFO data structure with a fixed, finite capacity. Using OCL, we can express class invariants to specify that the number of elements in a BoundedStack object must be greater than or equal to zero and less than or equal to the fixed capacity, as illustrated in Figure 2.

```
Class BoundedStack
Inv:  size() >= 0 (or equivalently self.size() >= 0)
      size() <= capacity()
```

**Figure 2: Class Invariants**

We can also express pre- and postconditions with respect to BoundedStack's methods, as illustrated in Figure 3. The `@pre` suffix is used to designate values prior to method invocation. The `result` keyword refers to the method's return value.

```
BoundedStack::push(Object obj):Object
Pre:  size() < capacity()
Post: size() = v@pre.size()+1 (where "=" denotes equality)
      v.lastElement() = obj
      result = v.lastElement()

BoundedStack::pop():Object
Pre:  not empty()  (where "not" denotes logical negation)
Post: result = v@pre.lastElement()
      size() = v@pre.size()-1

BoundedStack::peek():Object
Pre:  not empty()
Post: result = v.lastElement()
      v = v@pre

BoundedStack::empty():boolean
Post: result = (v.size() = 0)

BoundedStack::size():int
Post: result = v.size()

BoundedStack::capacity():int
Post: result = v.capacity()
```

**Figure 3: Pre- and Postconditions**

The foregoing example illustrates how OCL can be used to specify correctness constraints for Java programs. Specification, however, is only half of the equation. We also need the ability to validate such constraints at runtime. Unfortunately, OCL is silent on this issue. We must, therefore, look elsewhere for runtime support. This is where JDI comes in.

## 3   Java Debug Interface

JDI is a part of the new Java Platform Debugger Architecture ("JPDA"), which was released recently for use with the Java 2 SDK [13]. JPDA consists of a three-tiered API, including: (1) the Java Virtual Machine Debug Interface ("JVMDI"); (2) the Java Debug Wire Protocol ("JDWP"); and (3) JDI. JVMDI is a low level native interface that defines debugging services that must be provided/implemented by compliant VMs. JDWP is the protocol used for communication between a debugger and its target VM. Among other things, JDWP provides out-of-process and remote debugging capabilities. Finally, JDI provides high-level debugging features via a 100% pure Java interface that insulates users from various low-level platform dependencies.

Excluding exception classes, JDI defines only one class: Bootstrap. The rest of the API consists of interfaces. The Bootstrap class is used to obtain an instance of the VirtualMachineManager interface, which contains methods to manage connections to both local and remote target VMs. Three different mechanisms are provided for connecting a debugger to a target VM: (1) LaunchingConnectors enable debugger applications to launch a target VM session; (2) AttachingConnectors enable debuggers to attach to VMs that are already running; and (3) ListeningConnectors enable target VMs to attach to debuggers that are already running. In addition, JDI provides two interprocess communication mechanisms. Shared memory can be used for debuggers and target VMs running on the same machine. Sockets can be used for remote and distributed applications.

The JDI Connectors provide access to the target VM through an interface hierarchy rooted at the Mirror interface. The various subinterfaces of Mirror provide proxy style access to VM attributes. These subinterfaces provide the ability to trap target VM events, including method entry and exit events. In addition, they provide the ability examine and manipulate entities in the target VM. With standardized support for method entry and exit breakpoints as well as variable inspection and method invocation on target VM objects, JDI was a natural choice as our verification API.

## 4 Prototype

Using OCL and JDI, we have implemented a prototype correctness constraint auditor for Java programs. The basic architecture of the prototype consists of five principal components: (1) an auditor application; (2) an OCL parser / Java expression evaluator; (3) an input file consisting of OCL constraints; (4) an output file consisting of validation results; and (5) the target application. In a nutshell, the auditor application constructs a lookup table of OCL constraints from the input file and launches the target application in a separate VM with appropriate method entry and exit breakpoints enabled. Upon method entry and exit events, applicable invariants, preconditions and/or postconditions are retrieved from the lookup table and passed to the parser/evaluator for validation. The results of the validation are then written to the output file.

### 4.1 Auditor Application

The auditor application contains the main program logic. Using the parser/evaluator, the auditor constructs a lookup table of OCL constraints from the input file. Thereafter, the auditor launches the target application in a separate VM via JDI calls. Launching the target VM returns an instance of the VirtualMachine interface, which serves as a proxy for the composite state of the target VM within the auditor application. For purposes of our prototype, we configure monitoring for MethodEntryEvents and MethodExitEvents — two more JDI interfaces. Thereafter, the auditor uses an event loop to process the events that JDI deposits in an event queue represented by an instance of the EventSet interface.

MethodEntryEvents and MethodExitEvents cause the target VM to suspend execution. Accordingly, the auditor can process these events while the target VM is in a stable state. First, the auditor obtains an instance of interface Method, which provides access to an interrupted method's name, signature and declaring type. This information is used to retrieve lists of invariants, preconditions and/or postconditions from the lookup table. After retrieving the lists, the auditor iterates through them, passing individual constraints to the parser/evaluator for validation. Finally, the results of the evaluation are sent to the output file.

In accordance with Liskov's Substitution Principle, [14] if a derived class overrides a base class method: (1) the derived class preconditions must be equal to or weaker than the base class preconditions; and (2) the derived postconditions must be equal to or stronger than the base class postconditions. Within our prototype, as in Eiffel, this rule is implemented as follows: (1) derived invariants are combined with inherited invariants via the logical "and" operator; (2) derived preconditions are combined with inherited preconditions via the logical "or" operator; and (3) derived postconditions are combined with inherited postconditions via the logical "and" operator [6, 11, 15].

Although this approach adequately addresses the issues raised by Liskov's Substitution Principle, it has an undesirable side effect with respect to preconditions and dynamic binding. In the case of dynamic binding, the client's contract is with the static type of the invoking object. Accordingly, the preconditions associated with the static type are the ones that should be evaluated. By virtue of the logical "or" operation, however, a constraint will pass muster if the preconditions of either the static or the dynamic type are satisfied. Ultimately this does not effect runtime program correctness, because the runtime preconditions are properly evaluated. It does, however, undermine the *Design by Contract* methodology.

### 4.2 OCL Parser / Java Expression Evaluator

The parser/evaluator provides a number of different services. First, it provides parsing services for purposes of constructing the lookup table from the input file. In addition, it provides parsing and translation services for OCL expressions. Finally, it provides evaluation services for corresponding Java expressions. For parsing services, we utilized the JavaCC parser generator. [16] JavaCC generates top-down recursive descent parsers from supplied input grammars. We designed a custom grammar for the input file, and used the standard OCL grammar set forth in [12] to parse OCL expressions. Grammar-driven parser semantic actions translate OCL constructs to JDI method invocations. JDI runtime type checking verifies OCL type constraints, throwing exceptions on type mismatches.

To prevent side effects, we permit only side effect free methods to be used within OCL constraints. To prevent deadlock situations, we permit only non-synchronized methods to be used within OCL constraints. Both instance and static methods may be used within OCL constraints. In addition, we support the use of class attributes and method parameters using standard JDI calls. Support for the @*pre* suffix is provided by parsing postconditions to determine which @*pre* values will be needed. These values are then computed during MethodEntryEvent processing for subsequent use in connection with MethodExitEvents. Unfortunately, we are not able to support OCL's *result* keyword at this time, because the current release of JDI does not provide access to a method's return value within the context of a MethodExitEvent. This oversight will be remedied in an upcoming release of JDI.

## 5   Related Work

A number of tools for incorporating correctness constraints into Java have been proposed recently. The majority of these tools (*e.g.*, iContract, Jass, and AssertMate) specify constraints as comments in the original source code and use a preprocessor to convert the comments into code [17, 18, 19]. This approach, of course, requires access to the source code. In addition, it requires either the source code or the resulting class files to be modified, or both. As a result, this approach is not able to support correctness constraints for Java interfaces, which cannot contain executable code. Moreover, it results in two code bases: one with correctness constraints and one without. Administrative costs are, therefore, increased.

Another approach is represented by jContractor, which uses a design pattern framework and Java's reflective capabilities to add *Design by Contract* like assertions to Java programs [15]. The down side to this approach is that it requires programmer discipline, because each invariant, precondition and postcondition must be implemented as a distinct private boolean class method. In addition, this approach provides no mechanism for turning off constraint monitoring. To turn off monitoring under our approach, you simply run the program as you normally would since no changes are made to the class files.

In terms of functionality, Handshake comes closest to our approach [20]. Handshake uses a contract file containing constraint information and a dynamically linked library (DLL), which intercepts the VM's file operations in order to add constraint checking code on the fly at class loading time. Like our approach, Handshake does not require access to source code and requires no modifications to either the original source code or the Java VM. Unlike our approach, however, Handshake does not use an industry standard expression language and relies on a non-standard DLL to support runtime constraint evaluation. Furthermore, Handshake does not offer UML integration. Accordingly, we believe that our approach is superior.

## 6   Conclusion

Correctness constraints provide a foundation for automated debugging within object-oriented systems. Under the *Design by Contract* methodology, clients are responsible for ensuring preconditions and servers are responsible for guaranteeing postconditions and preserving invariants. Accordingly, when a precondition is violated, the bug may be attributed to client code. Similarly, when a postcondition or invariant is violated, the bug may be attributed to server code. With runtime assertion monitoring, the whole process of bug attribution becomes fully automated.

We have introduced a new approach to incorporating correctness constraints into Java development environments. Using OCL as our expression language and JDI as our verification API, correctness constraints are incorporated: (1) without any modifications to the Java language; (2) without any access or changes to existing source code; and (3) without any specialized VM services. As demonstrated by our prototype, this approach supports automated debugging of Java programs through precise specification and validation of Java programming constructs.